\documentclass[pre,twocolumn,superscriptaddress,amsfonts,amssymb,amsmath,showpacs]{revtex4-1}

\usepackage[dvips]{graphicx}
\usepackage{epsfig}

\usepackage{subfigure}

\begin{document}

\title{Micro-structure of damage in thermally activated fracture of
  Lennard-Jones systems}
\author{A.~Yamamoto$^1$, F.~Kun$^2$ and S.~Yukawa}

\affiliation{Department of Earth and Space Science, Osaka University,
  Toyonaka 560-0043, Japan \\ $^2$Department of Theoretical Physics,
  University of Debrecen, P.O.~Box 5, H-4010 Debrecen, Hungary}

\begin{abstract}
We investigate the effect of thermal fluctuations on the critical
stress and the micro-structure of damage preceding macroscopic
fracture of Lennard-Jones solids under a constant external load. 
Based on molecular dynamics simulations of notched specimens at finite
temperature, we show that the crystalline
structure gets distorted ahead of the crack in the secondary creep regime.
The damage profile characterizing the spatial distribution of lattice
distortions is well described by an exponential form.  
The characteristic length of the exponential form provides the
scale of damage which is found to be an increasing function of the
temperature: At low temperature damage is strongly localized to the
crack tip, while at high temperature damage extends to a broader
range leading to more efficient relaxation of overloads. As a
consequence, the stress intensity factor decreases with increasing 
temperature. The final macroscopic failure of the system occurs
suddenly which is initiated by the creation of vacancies and voids.
The creep strength exhibits the inverse square root scaling with the
notch size corrected by the extension of the process zone.
\end{abstract}
\maketitle

\section{Introduction}

Understanding sub-critical fracture occurring under constant or
periodic external 
loads below the fracture strength of materials is a very important
scientific problem with a broad spectrum of technological
applications. Depending on materials' details several microscopic
mechanisms may contribute to the time dependent response and final
failure of samples. In spite of this diversity, experiments and
theoretical investigations have revealed that sub-critical fracture
obeys scaling relations both on the macro and micro scales
\cite{alava1,alava2}.

Thermal activation of micro-crack nucleation and growth is one of the
primary mechanisms underlying sub-critical fracture. Theoretical and
experimental investigations have shown that both in homogeneous and
heterogeneous materials thermally driven stress fluctuations can be
responsible for the finite lifetime of loaded specimens
\cite{science_280_265_1998_gelfracture, ciliberto_1} leading to the
emergence of universal scaling laws such as the Andrade law and
time-to-failure power laws \cite{sornette_andrade_law,
  nechad_sornette_prl_2005}, furthermore, to the scaling behavior of
waiting times between micro-fractures \cite{ciliberto_2,
  cortet:205502,ciliberto_4,ciliberto_5,ciliberto_6}.  Experiments have also
revealed that thermally activated slow crack advancements affect even
the surface roughness of growing cracks \cite{mallick:255502}.

In order to understand the mechanisms of the sub-critical fracture of
crystalline and glassy materials at the atomistic scale, molecular
dynamics simulations \cite{allen} performed at finite temperatures
proved to be indispensable
\cite{thermo2,lj_fracture_2,lj_fracture_3,lj_fracture_4,lj_fracture_1,thermo1}.
Computer simulations of Lennard-Jones particle systems made possible
to understand the origin of the Griffith-type crack nucleation at the
atomic scale showing the importance of the creation of vacancy
clusters and voids
\cite{lj_fracture_2,lj_fracture_3,lj_fracture_4,lj_fracture_1,thermo1,yip_1,pugno,mattoni_prl_2005}.
The simulation technique was also extended to study dynamic fracture
phenomena \cite{marder_ijf_2004}.

In the present paper we study thermally activated sub-critical
fracture using molecular dynamics simulations of a Lennard-Jones
particle system in two dimensions. A rectangular sample with a
triangular lattice structure is constructed which is then subject to a
constant external load at different notch length varying also the
temperature. We show that the critical stress of the system has an
inverse square root dependence on the notch length as expected from
linear fracture mechanics (LFM) corrected by the extension of the
fracture process zone. The stress intensity factor at a given notch
length decreases with increasing temperature due to the more efficient
relaxation of overloads in the vicinity of the crack. In order to
characterize the micro-structure of the loaded sample, we introduce
the six-fold order parameter and analyze its spatial distribution
during the secondary creep regime. Computer simulations revealed that
damage concentrates ahead of the crack tip and decays to a nearly
homogeneous background at larger distances. The damage profile has an
exponential form from which the characteristic scale of damage can be
determined. Increasing the temperature the damage profile broadens and
the background damage increases which are consistent with the
decreasing stress intensity factor. The results are discussed in
comparison to the micro-structure of heterogeneous solids with only
quenched disorder under quasi-static loading just before ultimate
failure \cite{zapperi1}.
%\

\section{Model}

We perform molecular dynamics (MD) simulations of 
two dimensional solids which consists of particles. 
The particles interact through the Lennard-Jones (LJ)
potential
\begin{equation}
U(r)=4 \epsilon \left[ \left(\frac{a}{r} \right)^{12} -
  \left(\frac{a}{r} \right)^{6} \right],
\label{eq:ljpot}
\end{equation}
where $r, \epsilon$ and $a$ are the distance between particles, the strength of the potential and the characteristic length of the particle, respectively.  For the
computational efficiency, 
we cut the interaction potential Eq.~(\ref{eq:ljpot}) at the distance
$3a$ when calculating the inter-particle force, i.e.\ if the distance $r$
between two particles exceeds $3a$, the interaction disappears.
 The parameters $\epsilon$, $a$ and the mass of particles $m$
are set to be unity in all the simulations.  The temperature $T$ is
defined as the average kinetic energy of the system $T=
\sum_{i=1}^{N} m \mathbf{v}_{i}^{2} \slash 2 N k_{\mathrm{B}}$, where
$\mathbf{v}_{i}$ denotes the velocity of $i$-th particle, $N$ is the
number of particles, and $k_{\mathrm{B}}$ is the Boltzmann's constant.
Hereafter we also set the Boltzmann's constant $k_{\mathrm{B}}$ to be
unity.

\begin{figure}
\begin{center}
\epsfig{bbllx=0,bblly=140,bburx=580,bbury=700,
file=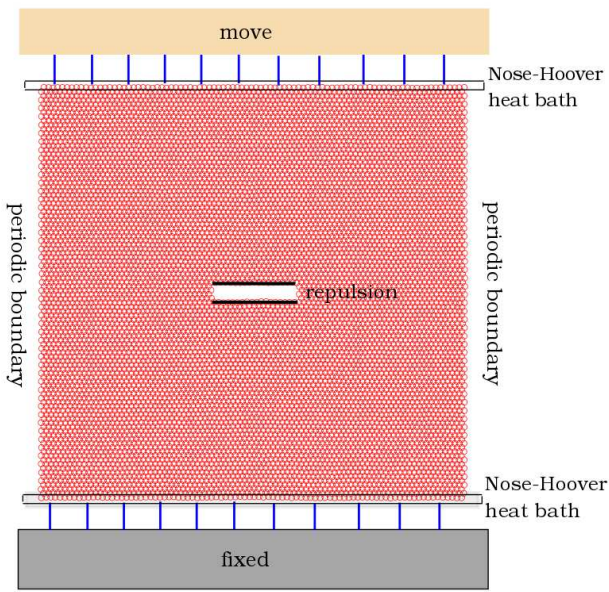, width=8.5cm}
\caption{{(Color online)} Schematic picture of the system: The
  sample is loaded through springs in the vertical direction.  Bottom
  and upper sides are attached to the Nose-Hoover thermostat to
  control the temperature. In the horizontal direction periodic
  boundary condition is applied.  }
\label{fig:setup}
\end{center}
\end{figure} 
A schematic picture of the computational geometry is presented in
Fig.~\ref{fig:setup}.  In order to control the temperature, 
Nose-Hoover thermostats are attached to two particle layers on the top
and bottom of the system \cite{nh1,nh2}.
In the horizontal direction we apply periodic boundary condition
to make the crack advance from the initial crack in the middle of the
system and to prevent the vaporization of particles due to the thermal
effect.

The initial state of the sample is prepared as follows: The particles
are placed on a regular triangular lattice at zero temperature $T=0$
then the temperature of the system is slowly increased up to the
desired value. We take the width of the system
as the corresponding equilibrium value for the temperature.  
Simulations were carried out with a fixed value of the width $l_0\simeq 100$ 
of the lattice.
As a crucial component of the model, an initial crack is created by removing four
particle layers with length $2L$ in the
middle of the system. In this way the crack has a nearly
rectangular shape with extensions $2L \times 3.5a$. Note that the distance $3.5a$
of the two sides  is larger than the cutoff length $3a$ of the potential,
which ensures that the removed particles form a crack in the model.
In order to prevent the crack from healing due to thermal fluctuation, the
particles on the upper and bottom crack faces interact through the
repulsive part of the LJ potential Eq.~(\ref{eq:ljpot}).  After
equilibrating the system, the load is applied in such a way that the
bottom and top particle layers are attached to loading bars through
spring elements. The bottom bar is fixed while the upper one can move
vertically controlling the load $\sigma$ on the system.  The bars are
treated as rigid bodies with a mass $10^{-3}$ per unit length, while
the loading springs have zero mass and a spring constant $10$.
The stiffness of the solid is about 83 for the parameters used in the
simulations, which is significantly larger than the one of springs.

In order to suppress the disturbing effect of elastic waves generated 
by the external loading process, the loading bar moves 
from the initial position until the desired stress value $\sigma$ is
reached at a relatively low increasing rate $d\sigma /dt =10^{-2}$.
Simulations 
showed that the value of the loading rate slower than $d\sigma /dt
=10^{-2}$ did not affect the time evolution of the system.

\section{Time evolution}

\begin{figure}
  \begin{center}
\epsfig{file=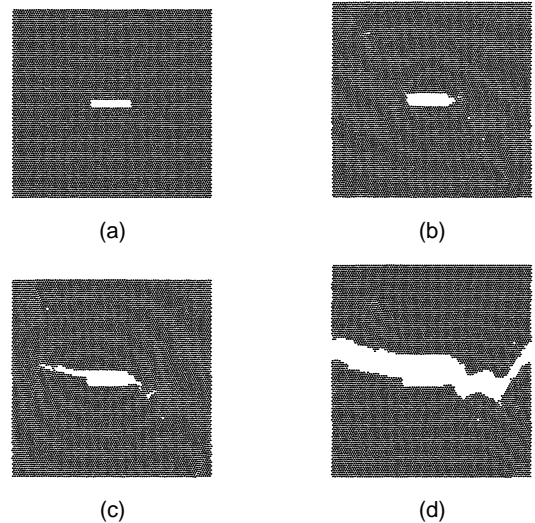,width=8.5cm}
  \end{center}
     \caption{Time evolutions of a sample at the 
       temperature $T=0.1$ and the load $\sigma=1.45$ 
       with the initial crack length $2L=20$: 
       (a) an initial state, and (b) a configuration after uploading and equilibration. 
       (c) As time evolves, the lattice structure gets distorted, the 
       crack grows and voids and defects nucleate ahead of the crack
       tip. 
       (d) The material's fracture occurs suddenly following a
       long evolution process during which the overall deformation
       hardly changes.    } 
\label{fig:time}
\end{figure}
The time evolution of the solid subject to a constant load is
calculated numerically by the second-order symplectic integrator method with the
time step $dt=10^{-3}$, while the time evolution of the thermostat and
the bar is obtained using the Euler method.  Figure~\ref{fig:time}
presents snapshots of the time evolution of the system at the
temperature $T=0.1$, the load $\sigma=1.45$, the initial crack length
$2L = 20$ and the width of the system $l_0 \simeq 100$. Starting from
the initial state (Fig.~\ref {fig:time}(a)) we gradually increase the
external load $\sigma$ up to the desired value and equilibrate the
system to achieve the desired temperature.  Figure~\ref{fig:time}(b)
shows the sample after the ramp loading, i.e., when the stress becomes
constant for the rest of the simulation.  It is observed in
Fig.~\ref{fig:time}(c) that the crystalline structure of the lattice
gets distorted.  Nucleation of defects and voids due to thermal
fluctuations is also observed, which facilitates the advancement of
the crack. When the specimen is separated into two parts by a 
crack like in Fig.~\ref{fig:time}(d), the system is considered to be fractured.

\begin{figure}
\begin{center}
\epsfig{file=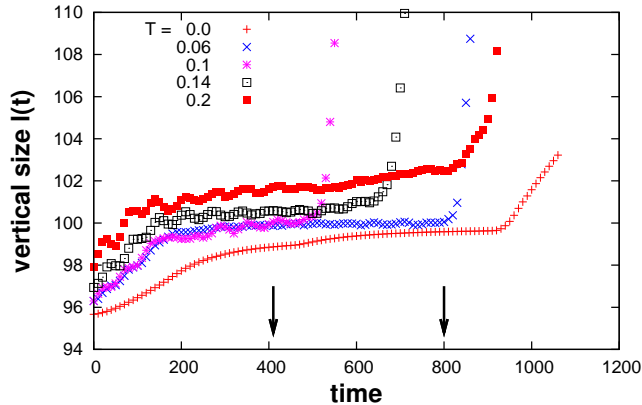, width=8.5cm}
     \caption{(Color online) Vertical size $l(t)$ of the system
       as a function of time for several temperatures with the initial
       crack length $2L=20$.  Elastic waves generated by the loading
       process and the void creations due to the thermal noise cause the small oscillations of $l(t)$.  The vertical
       arrows indicate the time window for the sample of $T=0.2$ where
       the damage parameter $d$ is evaluated (see
       Sec.~\ref{sec:damage}).  Applied loads are taken to be
       $\sigma=2.05, 1.7, 1.45, 1.35,$ and $\sigma=1.1$ for $T=0,
       0.06, 0.1, 0.14, $ and $T=0.2$, respectively.  }
\label{fig:def}
\end{center}
\end{figure}
On the macroscopic level, the time evolution of the system can be
monitored by measuring the vertical size $l(t)$ of the sample in the
direction of the external load as a function of time $t$.  This
quantity is presented in Fig.~\ref{fig:def} for several temperatures
with the initial crack length $2L=20$.
% For $T=0.0$ the uploading time is $t_{up} \sim 300$, while
% for all the finite temperature $T>0$ we set $t_{up} \sim 200$.
Note that the sample of Fig.~\ref{fig:time} corresponds to the curve
of $T=0.1$ in Fig.~\ref{fig:def}.  The plateau regime of the
deformation is the consequence of the slow dynamics of the system,
where the lattice structure gets gradually distorted while the length of the
crack hardly changes. The final macroscopic failure occurs rapidly,
which is characterized by the sudden acceleration of the vertical size
$l(t)$ and by its diverging derivative in Fig.~\ref{fig:def}.  The
acceleration of the vertical size is preceded by the creation of defects
and voids due to thermal activation.  Small oscillations are observed
along the plateau of $l(t)$, which are caused by elastic waves generated
by the loading process and the void creations due to the thermal noise.

\section{Macroscopic strength}

The main difficulty of the investigation of thermally activated
fracture of LJ systems is the enormous computational time required due
to the slow dynamics of the system. To overcome this problem, we
introduce a threshold time of the simulation; After the stress
$\sigma$ reaches the desired value, we limit the simulation time to
the finite threshold time $t_{th}$.  This threshold gives a lower
limit of the lifetime of the sample; If no complete failure is achieved
by following the time evolution of the system up to $t_{th}$, the
lifetime of the sample is assumed to be infinite at the given external
load $\sigma$.  The highest load at which no macroscopic failure
occurs before $t_{th}$ defines the critical load $\sigma_c$ of the
system which separates the regimes of finite and infinite lifetimes.
The critical load $\sigma_c$ of the system is determined 
by a sequence of simulations performed with a constant load increment 
$0.05$, which provides sufficient accuracy for $\sigma_c$. 
In this paper, the threshold time is fixed to be $t_{th}=10^3$.

\begin{figure}
\begin{center}
\includegraphics[width=25em]{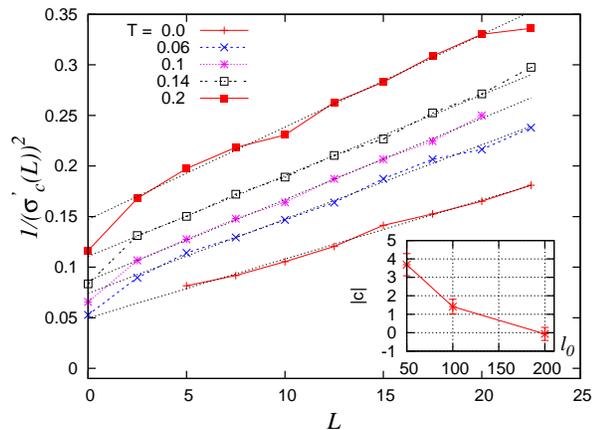}
\caption{(Color online) Crack length dependence of the critical
  stress: $\left( 1 \slash \sigma'_{c}(L) \right)^{2}$, where
  $\sigma'_{c}(L) =\sigma_{c}(L)-c$ and $c$ is a finite size
  correction, is plotted as a function of $L$ for several
  temperatures averaging over 10 samples.  The error bars
  are smaller than the symbol size. The dotted lines are obtained by
  fitting straight lines in a good agreement with Eq.~(\ref{eq:xifpz}).
  Deviations near $L>20$ of the data from the linear fit occur due to
  finite size effects, while the deviations for the small notch size
  indicate the dominance of thermal noise in the fracture process.
  Inset: System width $l_{0}$ dependence of the finite size correction
  $c$. }
\label{fig:ld}
\end{center}
\end{figure}

In order to investigate the effect of the notch size $2L$ and of the 
temperature on the macroscopic strength $\sigma_c$, we carry out
simulations with the following parameters:
%$2L=0, 5, 10, 15, 20, 25, 30, 35, 40,$ and $2L=45$ 
%and 
$2L=0, 5, 10, 15, 20, 25, 30, 35, 40,$ and $2L=45$ for the
temperatures $T=0, 0.06, 0.1, 0.14,$ and $T=0.2$.
We do not show the data of $2L=0, 5$ at $T=0$ because the crack did not grow at the crack tip.  
The value of $\sigma_c$ is determined by averaging over
10 samples at each parameter set.

Based on linear fracture mechanics we expect the inverse square
root dependence of the critical load $\sigma_c$ on the notch size $L$
as it has been proved by an analytical calculation \cite{bazant_1}.
In order to capture the disordering effect of temperature, we
introduce a characteristic length $\xi$ which modifies the notch size
into an effective one $L + \xi$.  Then we write $\sigma_c$ in the following
form 
\begin{equation}
\sigma_{c}(L,T) =  \frac{K}{\sqrt{L + \xi } }+ c,  
\label{eq:xifpz}
\end{equation}
where $K$ denotes the stress intensity factor \cite{zapperi1,bazant_1}
and $c$ is a finite size correction of the system width.  The finite
size correction $c$ is a negative value and its system size dependence
is plotted in the inset of Fig.~\ref{fig:ld}. It is important to emphasize that
the value of the correction factor $c$ goes to zero with increasing system
size $l_0$. 
To numerically demonstrate the validity of the functional 
form Eq.~(\ref{eq:xifpz}), in Fig.~\ref{fig:ld} we plot $( 1\slash
\sigma^{'}_{c}(L) )^{2}$, where $\sigma^{'}_{c}(L)=\sigma_{c}(L) -c$,
as a function of half of the notch length $L$.  It is observed in
Fig.~\ref{fig:ld} that straight lines are obtained with a good
accuracy at all temperatures by fitting.  Deviations from the linear
form are observed for both large and small values of the initial crack
length: In the large crack length regime $20\lesssim L$, deviations
occur due to the finite size of the system,
on the other hand, in the limit of small cracks $L\lesssim 2.5$, the
stress concentration at the crack tip becomes less dominating so that
the fracture process is mainly controlled by thermal noise emerging in
the entire volume of the system.  By fitting these data from $L=2.5$
to $20$, we estimate the stress intensity factor $K$ and the
characteristic length $\xi$ of Eq.~(\ref{eq:xifpz}) for each
temperature.

\begin{figure}
\begin{center}
\mbox{
\subfigure[]{\epsfig{file=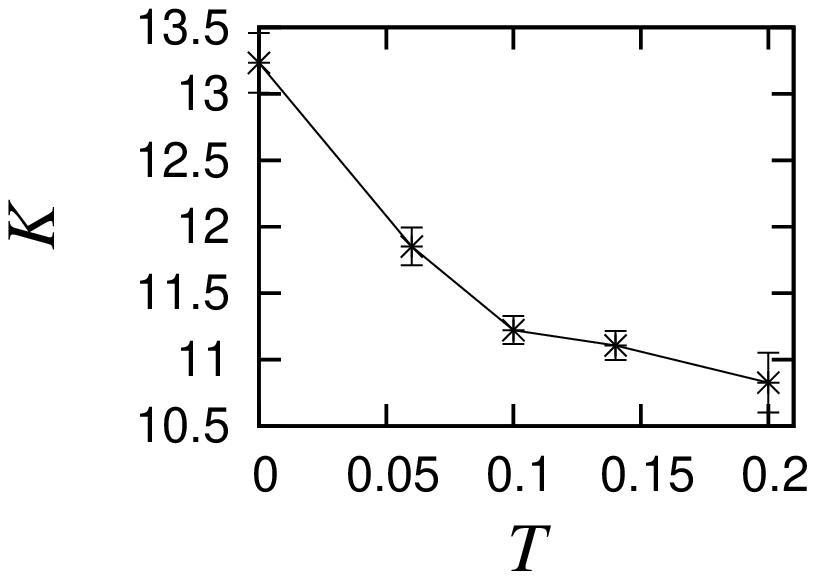,angle=270,width=4.3cm}}
\subfigure[]{\epsfig{file=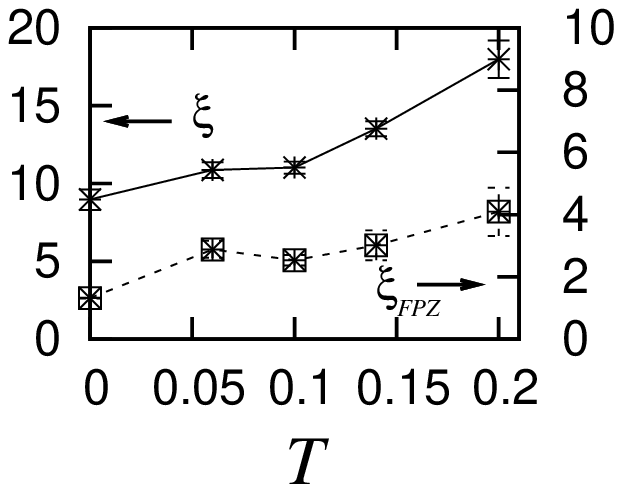,angle=270,width=4.3cm}}}
\caption{(a) Temperature dependence of the stress intensity factor $K$: As
  the temperature increases the value of $K$ is monotonically decreasing. 
  (b) Solid line shows the temperature dependence of
  the characteristic length $\xi$ defined by Eq.~(\ref{eq:xifpz}). 
  Dashed line represents the extension of the distorted zone 
 $\xi_{FPZ}$ determined by the analysis of the micro-structure of the
 system (This analysis is carried out in Sec.~\ref{sec:damage}.).
 Simulations are carried out with a fixed notch size $2L=20$.} 
\label{fig:stin}
\end{center}
\end{figure}
Figs.~\ref{fig:stin}(a) and (b) present the temperature dependence of
the stress intensity factor $K$ and of $\xi$, respectively.  It is
observed that the stress intensity factor $K$ decreases with
increasing temperature $T$, while the value of $\xi$ is an increasing
function of $T$. In Fig.~\ref{fig:stin}(a), at higher temperature thermally
activated particle motion has a higher intensity which facilitates the
relaxation of the system and gives rise to a decreasing $K$.  In
Fig.~\ref{fig:stin}(b) the results imply that at higher temperature 
a broader zone is formed ahead of the crack tip where 
distortion of the crystalline structure and plastic deformation can occur 
due to the interplay between the stress field and the
thermal noise. The length scale $\xi$ can be interpreted as the
linear
extension of the fracture process zone (FPZ) ahead of the crack tip
which appears due to the disordering effect of thermal noise on the
lattice structure. In the vicinity of the crack tip the high stress
concentration makes the system more sensitive to thermal
fluctuations. As a consequence, a distorted zone emerges which affects
the scaling of $\sigma_c$ with the notch size $L$. Similar effect is
observed in heterogeneous quasi-brittle materials where damage in the
form of micro-cracks is concentrated ahead of the crack
\cite{zapperi1,bazant_1}.

\section{Micro-structure of damage}
\label{sec:damage}

\begin{figure}
\begin{center}
\includegraphics[width=7cm]{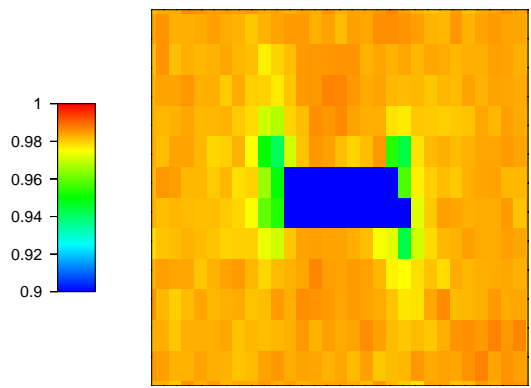}
\end{center}
     \caption{(Color online) Spatial distribution of the
       bond-orientational order parameter averaged on a rectangular $2
       \times 2 $ mesh: The temperature and the initial crack length
       are taken to be $T=0.06$ and $2L=20$. For the sake of clarity,
       we present a magnified view of the vicinity of the crack.  The
       color code indicates the modification of the triangular-crystalline
       structure due to the high stress concentration and to the
       thermal noise.  Deep blue indicates the order parameter less
       than $0.9$.  In this figure, the blue region corresponds to the
       crack.  }
\label{fig:order}
\end{figure}
In order to obtain a direct quantitative measure of the extension of
the process zone formed ahead of the crack tip, we carry out a
detailed analysis of the micro-structure of the system before the
acceleration of deformation starts.  To characterize the crystalline
order, we introduce the bond-orientational order parameter $\phi_k$
defined for particle $k$ as follows \cite{order_param}
\begin{equation}
\phi_{k} =   \left| \sum_{l}^{n_{k}} \frac{
    \exp(6i\theta_{kl})}{n_{k}} \right|.
\end{equation}
Here $n_{k}$ denotes the number of neighbors of particle $k$ and
$\theta_{kl}$ is an angle between a fixed axis such as the $x$ axis
and the bond connecting particle $k$ to particle $l$.
The value of $\phi_k$ can vary between zero and one such that
$\phi_k=1$ quantifies a perfectly ordered triangular lattice, while
$\phi_k\approx 0$ is obtained for a disordered fluid-like system.
Intermediate values of the order parameter characterize the degree of
distortion of the lattice structure.

In order to characterize the spatial variation of the
micro-structure, we calculate a locally averaged bond-orientational
order parameter $\phi$ on a coarse-grained mesh in the sample.  The
mesh size is taken to be $2 \times 2$.
%This procedure smoothens the order parameter. 
%allowing us to deduce information on its spatial variation. 
The order parameter is evaluated for a single sample, but $30$
snapshots are taken from the plateau regime of the deformation-time
diagrams in order to improve the statistics.  As shown in
Fig.~\ref{fig:def}, the two vertical arrows indicate the time window
of the plateau regime for the case of temperature $T=0.2$.
%where 30 snapshots are averaged with the
%meshing techniques. 
A typical example of the order parameter $\phi$ for the temperature
$T=0.06$ is presented in Fig.~\ref{fig:order}.  It is observed that
the crystalline structure has strong distortion in the vicinity of the
crack tip where the stress concentration is high.  Away from the crack
tip, and even near the middle of the crack, the order parameter is
homogeneously distributed and it has a large value close to unity,
i.e.~the crystalline structure is retained there.

\begin{figure}
\begin{center}
\includegraphics[width=25em]{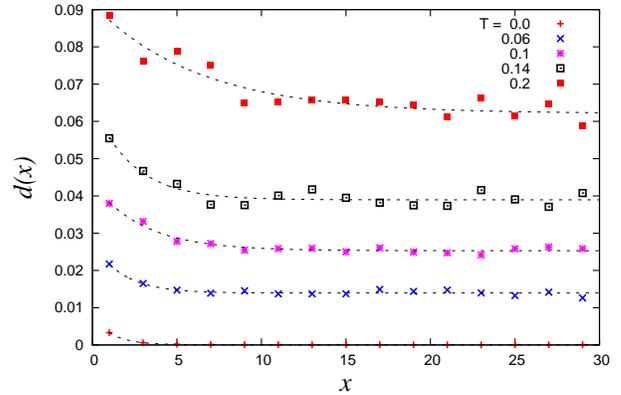}
\end{center}
     \caption{(Color online) Damage profile, i.e. the damage parameter
       $d$ as the function of the distance $x$ from the crack tip for
       the initial crack length $2L=20$ at several temperatures.  The damage
       profile exhibits an exponential decay with the form
       Eq.~(\ref{eq:exponential}) to a uniform background damage. All
       the three parameters characterizing the functional form of
       $d(x)$ depend on the temperature (see Fig.~\ref{fig:stin}(b) and Fig.~\ref{fig:dampar}). }
\label{fig:dam}
\end{figure}
\begin{figure}%[!h]
\begin{center}
\epsfig{file=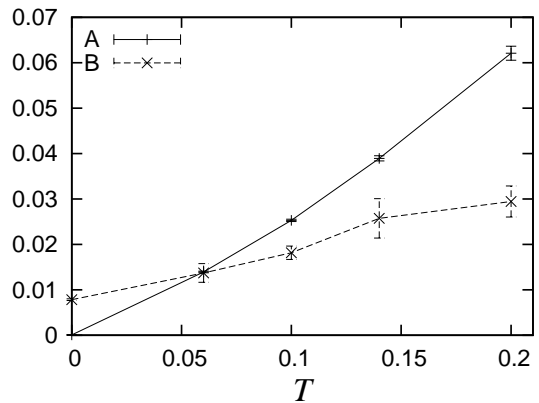, width=8cm}
\end{center}
     \caption{The parameters $A$ and $B$ of the damage profile as
       function of temperature for the notch size $2L=20$. The value
       of the background disorder $A$ rapidly increases with the
       temperature, while the multiplication factor $B$ has a moderate
       increase. }
\label{fig:dampar}
\end{figure}
To obtain a detailed quantitative measure of the disordering effect of
the stress concentration assisted by the thermal noise, we introduce a
damage parameter with the following definition
\begin{eqnarray}
d_k = 1-\phi_k,
\end{eqnarray}
which has a large value $d_k \approx 1$ at locations where the lattice
structure is most distorted.  Based on the above meshing technique,
we average $d_k$ on each plaquette of the mesh, then make a projection on the
horizontal axis, and determine the damage parameter along the initial
crack axis as a function of the distance from the crack tip $x$; At
each position $x$, the averaged damage parameter is averaged again on
the vertical column of 5 plaquettes.  The damage profile $d(x)$ obtained
by this way is presented in Fig.~\ref{fig:dam} for several different
temperatures.  Note that the calculations are performed at the same
load value $\sigma =1.0$
%which is above the critical load 
for all the temperatures at the given notch length $2L=20$. 
It is observed that $d(x)$ monotonically decreases with increasing
distance $x$ from the crack tip as it is expected, and converges to
finite values at long distances. The figure presents that the profiles
can be well described by the exponential form
\begin{eqnarray}
\label{eq:exponential}
d(x) = A+B \exp \left(-x\slash\xi_{FPZ} \right ),
\end{eqnarray}
where the parameter $A$ represents the uniform background damage,
$\xi_{FPZ}$ is the characteristic length of the extension of the
distorted zone, and the multiplication factor $B$ represents the
maximum value of the additional damage above the background achieved
at the crack tip.  The functional form Eq.~(\ref{eq:exponential})
provides an excellent fit of the numerical data in all
cases. Therefore we determine the temperature dependence of the
parameters $A$, $B$, and $\xi_{FPZ}$ (see Fig.~\ref{fig:dampar} and
Fig.~\ref{fig:stin}(b)).
%(see Fig.~\ref{fig:dampar}). 
Note that the damage parameter $d$ at zero temperature decreases to
zero at large distances, while it takes non-zero value at the crack
tip.  This clearly indicates that the distortion of the lattice is
caused by the high stress concentration arising in the vicinity of the
crack tip.
%Hence, from the fitting
%$A=0$ and a relatively large $B$ and small $\xi_{FPZ}$ values are obtained
%at $T=0.0$ (see Fig.~\ref{fig:stin}(b) and Fig.~\ref{fig:dampar}).
Increasing the temperature, the thermal noise gets stronger which
lightens the stress concentration near the crack tip by distorting the
lattice structure.  This mechanism has the consequence that with
increasing temperature the localization of stress at the crack tip
decreases which is quantified by the increasing characteristic length
$\xi_{FPZ}$ (see Fig.~\ref{fig:stin}(b) and Fig.~\ref{fig:dampar}).

The most important characteristics of the micro-structure is
represented by the length scale $\xi_{FPZ}$ which provides a measure
of the extension of the highly distorted zone. It can be observed in
Fig.~\ref{fig:stin}(b) that $\xi_{FPZ}$ increases when the
temperature gets higher. It shows that increasing thermal noise in the
system gives rise to a larger extension of the fracture process zone.
It can also be observed in Fig.~\ref{fig:stin}(b) that $\xi$ and
$\xi_{FPZ}$ have a linear relation which implies that the
micro-structure based length $\xi_{FPZ}$ provides a good measure of
the macroscopic fracture process zone.

\section{Conclusions and Discussion}

We carried out a detailed study of the creep rupture of Lennard-Jones
particle systems by means of molecular dynamics simulations. In order
to control the temperature a Nose-Hoover thermostat was coupled to the
sample which was then subject to a constant load through spring
elements. The main focus of our work was to understand the effect of
thermal fluctuations on the creep strength and to investigate the
micro-structure of the evolving system before the onset of rapid crack
growth leading eventually to macroscopic failure. Simulations showed
that the critical stress of samples has an inverse square root
dependence on the notch length corrected by the extension of the
fracture process zone. The fracture process zone arises due to the
distortion of the lattice structure as a consequence of the interplay
of the high stress concentration at the crack tip and of the thermal
noise. The final failure occurs rapidly after the creation of
vacancies and voids in the process zone.

In order to obtain a direct quantitative measure of the extension of
the fracture process zone, we introduced a damage parameter and analyzed
its spatial distribution. Our calculations revealed that damage
concentrates in the vicinity of the crack tip and decays exponentially
to a homogeneous background damage level. In spite of the inverse
square root form of the stress distribution ahead of the crack expected
from linear fracture mechanics, the damage profile is described by an
exponential form which let us introduce a characteristic length scale
of damage $\xi_{FPZ}$. The highly distorted regime of extension
$\xi_{FPZ}$ is analogous to the fracture process zone observed in
various types of systems.

A similar exponential form of the damage profile was recently found
for heterogeneous materials where thermal activation does not play a
role \cite{zapperi1}. The authors investigated quasi-brittle fracture
processes in the framework of the fuse model gradually increasing the
load on notched samples up to failure. Evaluating the density of
micro-cracks in the critical state of the sample just before
macroscopic failure, it was found that damage concentrates ahead of the
crack. The damage profile proved to have an exponential form similar
to our one. It was argued that micro-cracks concentrate at the crack
tip due to the high stress concentration, however, as a counter effect
they screen the stress field which then leads to the exponential decay
instead of the inverse square root form \cite{zapperi1}.

A very interesting question addressed by Ref.~\onlinecite{zapperi1} is
the crossover between the two regimes where the fracture strength of
the system is dominated by stress concentration at the notch and by
the disorder, respectively. The deviation from the linear behavior in
Fig.~\ref{fig:ld} at the limit of small notch sizes already indicates
the dominance of thermal noise in the fracture process. However, due
to technical reasons we could not further decrease the initial crack
length which prevents us to deduce quantitative conclusion on the
crossover phenomenon.

\begin{acknowledgments}
We acknowledge support of the MTA-JSPS project 32/2008, the TeT project
JP-24/09, the Complexity-Net, and a Grant-in-Aid for Scientific Research (C) No.
22540387, from the Ministry Education, Culture, Sports, Science and Technology
Japan. The work is supported by TAMOP-4.2.1/B-09/1/KONV-2010-0007 project. The
project is implemented through the New Hungary Development Plan,
co-financed by the European Social Fund and the European Regional
Development Fund.
F.\ Kun acknowledges support of the Janos Bolyai
project of HAS and of OTKA K84157.  A.\ Yamamoto acknowledges support of 
Global COE Program (Core Research and Engineering of Advanced 
Materials-Interdisciplinary Education Center for Materials Science), MEXT, Japan and the JSPS 
International Training Program(ITP).

\end{acknowledgments}

\end{document}